\documentclass[12pt]{iopart}
\usepackage{hyperref}
\usepackage{graphicx}

\begin{document}
\title[Electrostatic potential of a uniformly charged disk]
{On the electrostatic potential and electric field of a uniformly charged disk}
\author{V Bochko $^1$,  Z K Silagadze$^{1,2}$}
\address{$^1$ Novosibirsk State University, Novosibirsk 630090, Russia}
\ead{v.bochko@g.nsu.ru}
\address{$^2$ Budker Institute of
Nuclear Physics, Novosibirsk 630 090, Russia}
\ead{Z.K.Silagadze@inp.nsk.su}



\begin{abstract}
We calculate the electrostatic potential and electric field of a uniformly 
charged disk everywhere in space. This electrostatic problem was solved long
ago, and its gravitational analogue --- even earlier. However, it seems that
physics students are not aware of the solution, because it is not presented 
in textbooks. The purpose of the present article is to fill this gap in the 
pedagogical literature. 
\end{abstract}
\submitto{\EJP}
\maketitle

\section{Introduction}  
The problem of finding the electric field on the  axis of symmetry of 
a uniformly charged circular disk is considered in many introductory textbooks 
on classical electrodynamics (see, for example, \cite{1,2,3,4,4A,5}). 
Naturally, students may wonder how the electric field (or the electrostatic 
potential) can be found in a more general case for points that do not lie on 
the axis of symmetry. Surprisingly, students will not find the answer even in 
more advanced textbooks \cite{6,6A,6B,7,8,9,9A,9B,9C,9D}. In \cite{9} a hint 
is given without any further details that the electric field around 
a homogeneously charged  disk generally requires expression in terms of 
elliptic integrals. 

The desired  general answer in terms of elliptic integrals can be found in 
a comprehensive French treatise \cite{10}, in Russian textbooks on potential 
theory \cite{11,12}, or in special scientific literature \cite{13,14,15}.
However, we are afraid that not one of them are especially accessible to 
the average student.

In an interesting exercise 2.53 in \cite{5} (marked as a difficult four-star
problem),  it is proposed to show that the radial component of the electric 
field of a uniformly charged disk, at a point $P$ located at a distance 
$\eta R$ from the center of the disk (where $0\le\eta<1$ and $R$ is the radius 
of the disk) and at an infinitesimal distance away from the plane of the disk,
has the following form 
\begin{equation}
E_r=\frac{\sigma}{2\pi\epsilon_0}\int\limits_0^{\pi/2}\ln{\left (\frac{
\sqrt{1-\eta^2\sin^2{\theta}}+\eta\sin{\theta}}{\sqrt{1-\eta^2\sin^2{\theta}}
-\eta\sin{\theta}}\right)}\cos{\theta}d\theta,
\label{eq1}
\end{equation}
where $\sigma$ is surface charge density on the disk. 

\begin{figure}[htp]
\centering
\includegraphics[height=4cm]{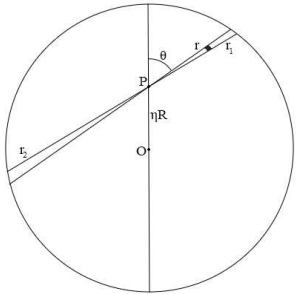}
\caption{Dividing the disk into infinitely small wedges centered at point P.
The black  patch at  distance $r$ from $P$ is a wedge element containing  
the charge $dq=\sigma r drd\theta$.}
\label{fig1}
\end{figure}

Instead of using cylindrical coordinates associated with the center of the 
disk, which might seem natural due to the symmetry of the problem, a hint in 
the book suggests to divide the disk into infinitely small wedges centered at 
point P, as shown in Fig.\ref{fig1}, then find  the  sum  of  the  fields from 
the two opposite wedges, and finally integrate over the angle $\theta$.

This clever trick can already be found in \cite{16}, and it was again 
discovered in the form of the so-called strip function technique in \cite{17}
when studying the electrostatics and magnetostatics of a conducting disk. 
This method (without much emphasis) was used in \cite{13} when calculating
the gravitational potential due to uniform disk. 

The purpose of this note is to demonstrate that with a certain level of 
moderate mathematical complexity (basic knowledge of elliptic integrals), 
the same technique allows to find the electrostatic potential and electric 
field of a uniformly charged disk everywhere in space, and not just on 
the disk.

\section{Electrostatic potential on the disk}
\begin{figure}[htp]
\centering
\includegraphics[height=4cm]{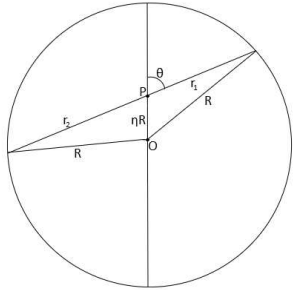}
\caption{Illustration of the calculation of  $r_1(\theta)$ and  $r_2(\theta)$
from the law of cosines.}
\label{fig2}
\end{figure}

An element of the right wedge in Fig.\ref{fig1}, at distance $r$ from $P$,
contains an amount of charge $dq=\sigma r drd\theta$ and, thus, creates at $P$
the electrostatic potential $dq/(4\pi\epsilon_0 r)= \sigma drd\theta/(4\pi
\epsilon_0)$. The contribution of the entire right wedge is then
\begin{equation}
d\phi_R(P)=\frac{\sigma d\theta}{4\pi \epsilon_0}\int\limits_0^{r_1}dr=
\frac{\sigma r_1}{4\pi \epsilon_0}d\theta.
\label{eq2}
\end{equation}
The same consideration applies to the left wedge with the result
\begin{equation}
d\phi_L(P)=\frac{\sigma r_2}{4\pi \epsilon_0}d\theta.
\label{eq3}
\end{equation}
When $\theta$ changes from $0$ to $\pi/2$, a half of the disk's area is
covered. It is evident from the symmetry that another half gives the same
contribution to the potential at $P$. Therefore, finally
\begin{equation}
\phi(P)=\frac{\sigma}{2\pi \epsilon_0}\int\limits_0^{\pi/2}\left [ r_1(\theta)+
r_2(\theta)\right ]d\theta.
\label{eq4}
\end{equation}
The functions $r_1(\theta)$ and  $r_2(\theta)$ can be found from
Fig.\ref{fig2}.
Namely, using  the law of cosines, we get
\begin{equation}
R^2=r_1^2+\eta^2R^2+2\eta r_1R\cos{\theta}=
r_2^2+\eta^2R^2-2\eta r_2R\cos{\theta}.
\label{eq5}
\end{equation}
Solving quadratic equations (\ref{eq5}) for $r_1$ and $r_2$, we get (note that
$\eta\cos{\theta}<\sqrt{1-\eta^2\sin^2{\theta}}$, since $\eta<1$)
\begin{equation}
r_1=R\left[\sqrt{1-\eta^2\sin^2{\theta}}-\eta\cos{\theta}\right],\;\;
r_2=R\left[\sqrt{1-\eta^2\sin^2{\theta}}+\eta\cos{\theta}\right].
\label{eq6}
\end{equation}
Therefore, (\ref{eq4}) takes the form
\begin{equation}
\phi(P)=\frac{\sigma R}{\pi\epsilon_0}\int\limits_0^{\pi/2}\sqrt{1-
\eta^2\sin^2{\theta}} d\theta=\frac{\sigma R}{\pi\epsilon_0}\,E(\eta),
\label{eq7}
\end{equation}
where
\begin{equation}
E(k)=\int\limits_0^{\pi/2}\sqrt{1-k^2\sin^2{\theta}} d\theta
\label{eq8}
\end{equation}
is the complete elliptic integral of the second kind \cite{18,18A}.

For the radial component of the electric field at the point $P$ we get from
(\ref{eq7}) the following expression
\begin{equation}
E_r(P)=-\frac{1}{R}\,\frac{\partial \phi_P}{\partial \eta}=\frac{\sigma\eta}
{\pi\epsilon_0}\int\limits_0^{\pi/2}\frac{\sin^2{\theta}}{\sqrt{1-
\eta^2\sin^2{\theta}}}\,d\theta=\frac{\sigma}{\pi\epsilon_0}\,
\frac{K(\eta)-E(\eta)}{\eta},
\label{eq9}
\end{equation}
where 
\begin{equation}
K(k)=\int\limits_0^{\pi/2}\frac{d\theta}{\sqrt{1-k^2\sin^2{\theta}}}
\label{eq10}
\end{equation}
is the complete elliptic integral of the first kind and at the last step we 
have used the relation \cite{18,18A}
\begin{equation}
\frac{dE(k)}{dk}=\frac{E(k)-K(k)}{k}.
\label{eq11}
\end{equation}
At first sight (\ref{eq1}) and (\ref{eq9}) look different. However, we can
write (\ref{eq1}) in the form
\begin{equation}
E_r=\frac{\sigma}{2\pi\epsilon_0}\left [I(\eta)-I(-\eta)\right ],
\label{eq12}
\end{equation}
where the integral $I(\eta)$ can be transformed using integration by parts
as follows:
\begin{equation}
\fl I(\eta)=\int\limits_0^{\pi/2}\ln{[\sqrt{1-\eta^2\sin^2{\theta}}+\eta
\cos{\theta}]}\,d\sin{\theta}=\ln{\sqrt{1-\eta^2}}+\int\limits_0^{\pi/2}
\frac{\eta\sin^2{\theta}}{\sqrt{1-\eta^2\sin^2{\theta}}}\,d\theta, 
\label{eq13}
\end{equation}
and this makes clear that (\ref{eq1}) and (\ref{eq9}) are, in fact, equivalent.

\section{Electrostatic potential above or below the disk}
\begin{figure}[htp]
\centering
\includegraphics[height=4cm]{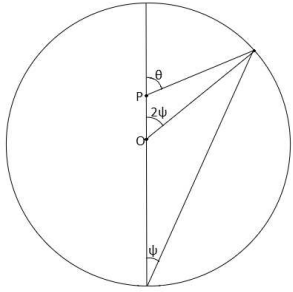}
\caption{The meaning of the angle $\psi$.}
\label{fig3}
\end{figure}

Now let's assume that the point $P$ in Fig.\ref{fig1} is just the projection
on the disk of the point where we want to find the electrostatic potential.
Then instead of (\ref{eq2}) and (\ref{eq3}) we will have 
\begin{equation}
d\phi_R=\frac{\sigma d\theta}{4\pi \epsilon_0}\int\limits_0^{r_1}\frac{r}
{\sqrt{r^2+z^2}}\,dr=\frac{\sigma d\theta }{4\pi \epsilon_0}\left (\sqrt{r_1^2+
z^2}-|z|\right ),
\label{eq14}
\end{equation}
and 
\begin{equation}
d\phi_L=\frac{\sigma d\theta}{4\pi \epsilon_0}\int\limits_0^{r_2}\frac{r}
{\sqrt{r^2+z^2}}\,dr=\frac{\sigma d\theta }{4\pi \epsilon_0}\left (\sqrt{r_2^2+
z^2}-|z|\right ),
\label{eq15}
\end{equation}
Therefore, in this case the electrostatic potential takes the form
\begin{equation}
\phi(\eta R,z)=\frac{\sigma}{2\pi \epsilon_0}\int\limits_0^{\pi/2}\left (
\sqrt{r_1^2+z^2}+\sqrt{r_2^2+z^2}-2|z|\right )d\theta,
\label{eq16}
\end{equation}
where $r_1(\theta)$ and $r_2(\theta)$ are given by (\ref{eq6}) and it is clear
from these expressions that $r_2(\pi-\theta)=r_1(\theta)$. Then
\begin{equation}
\int\limits_0^{\pi/2}\sqrt{r_2^2+z^2}\,d\theta=\int\limits_{\pi/2}^\pi
\sqrt{r_1^2+z^2}\,d\theta,
\label{eq17}
\end{equation}
and (\ref{eq16}) can be rewritten as 
\begin{equation}
\phi(\eta R,z)=\frac{\sigma}{2\pi \epsilon_0}\left (\int\limits_0^{\pi}
\sqrt{r_1^2+z^2}\,d\theta-\pi|z|\right ).
\label{eq18}
\end{equation}
To calculate the integral in (\ref{eq18}), it is convenient to introduce
the angle $\psi$ shown in Fig.\ref{fig3} \cite{16} and use  $\phi=\pi/2-\psi$
as the new integration variable instead of $\theta$.

Since $OP=\eta R$, it is clear from the figure Fig.\ref{fig3} that
\begin{equation}
\tan{\theta}=\frac{R\sin{2\psi}}{R\cos{2\psi}-\eta R}=
\frac{\sin{2\psi}}{\cos{2\psi}-\eta},
\label{eq19}
\end{equation}
and therefore
\begin{equation}
d\theta=2\,\frac{1-\eta{\cos{2\psi}}}{1+\eta^2-2\eta\cos{2\psi}}\,d\psi=
-2\,\frac{1+\eta{\cos{2\phi}}}{1+\eta^2+2\eta\cos{2\phi}}\,d\phi.
\label{eq20}
\end{equation}
Fig.\ref{fig3} shows that when $\theta$ changes from $0$ to $\pi$, $\psi$
changes from $0$ to $\pi/2$ and, correspondingly, $\phi$ changes from $\pi/2$
to $0$. Therefore, the integral in (\ref{eq18}) can be written in the 
following way: 
\begin{equation}
\fl \int\limits_0^{\pi}\sqrt{r_1^2+z^2}\,d\theta=\frac{2R\sqrt{(1+\eta)^2+
z^2/R^2}}{(1+\eta)^2}\int\limits_0^{\pi/2}\frac{(1+\eta-2\eta\sin^2{\phi})
(1-k^2\sin^2{\phi})}{(1-n^2\sin^2{\phi})\sqrt{1-k^2\sin^2{\phi}}}\,d\phi.
\label{eq21}
\end{equation}
Here we have used $r_1^2=R^2(1+\eta^2-2\eta\cos{2\psi})=R^2(1+\eta^2+
2\eta\cos{2\phi})$ (from Fig.\ref{fig3}), $\cos{2\phi}=1-2\sin^2{\phi}$ and 
introduced notations 
\cite{13}
\begin{equation}
n^2=\frac{4\eta}{(1+\eta)^2}<1,\;\;k^2=\frac{4\eta}{(1+\eta)^2+z^2/R^2}<1,
\label{eq22}
\end{equation}
which allow to write
\begin{equation}
\eqalign{1+\eta^2+2\eta\cos{2\phi}=(1+\eta)^2(1-n^2\sin^2{\phi}), \cr 
r_1^2+z^2=R^2\left[(1+\eta)^2+\frac{z^2}{R^2}\right ]
\left(1-k^2\sin^2{\phi}\right).}
\label{eq23}
\end{equation}
To express (\ref{eq21}) in terms of complete elliptic integrals, let's 
decompose
\begin{equation}
\fl \frac{(1+\eta-2\eta\sin^2{\phi})(1-k^2\sin^2{\phi})}{1-n^2\sin^2{\phi}}=
A+B(1-k^2\sin^2{\phi})+\frac{C}{1-n^2\sin^2{\phi}}.
\label{eq24}
\end{equation}
For unknown coefficients $A,B,C$ we get a system of equations
\begin{equation}
\eqalign{Bn^2=2\eta,\cr A+B+C=1+\eta,\cr An^2+B(n^2+k^2)=2\eta+(1+\eta)k^2.} 
\label{eq25}
\end{equation}
The solution of this system is easily found to be
\begin{equation}
A=\frac{k^2}{2n^2}\,(1-\eta^2),\;\;B=\frac{2\eta}{n^2}=\frac{(1+\eta)^2}{2},
\;\; C=\frac{1}{2}(1-\eta^2)\left [1-\frac{k^2}{n^2}\right].
\label{eq26}
\end{equation}
Given (\ref{eq21}), (\ref{eq22}), (\ref{eq24}) and (\ref{eq26}),
the electrostatic potential (\ref{eq18}) can be expressed in terms of complete 
elliptic integrals as follows (in full agreement with the results of 
\cite{13}):
\begin{equation}
\fl\eqalign{
\;\;\;\;\;\phi(\eta R,z)=\frac{\sigma R}{2\pi\epsilon_0}\left[\frac{1-\eta^2}
{\sqrt{(1+\eta)^2+z^2/R^2}}\,K(k)+\sqrt{(1+\eta)^2+z^2/R^2}\,E(k)+\right .\cr
\left . \;\;\;\;\;
\frac{1-\eta}{1+\eta}\frac{z^2/R^2}{\sqrt{(1+\eta)^2+z^2/R^2}}\,\Pi(n^2,k)-
\pi\,\frac{|z|}{R}\right].}
\label{eq27}
\end{equation}
Here 
\begin{equation}
\Pi(n^2,k)=\int\limits_0^{\pi/2}\frac{d\theta}{(1-n^2\sin^2{\phi})
\sqrt{1-k^2\sin^2{\theta}}}
\label{eq28}
\end{equation}
is the complete elliptic integral of the third kind \cite{18,18A}.

If the observation point $P$ is situated on the $z$-axis, then $\eta=0$,
$k^2=n^2=0$, and using $K(0)=E(0)=\Pi(0,0)=\pi/2$, (\ref{eq27}) simplifies
to the well-known expression
\begin{equation}
\phi(0,z)=\frac{\sigma }{2\pi\epsilon_0}\left (\sqrt{z^2+R^2}-|z|\right).
\label{eq28A}
\end{equation}

\section{Electric field above or below the disk}
Due to rather complicated character of the expression (\ref{eq27}) for the
electrostatic potential, it will be easier to calculate the components of
the corresponding electric field using (\ref{eq18}) instead of the 
integrated form (\ref{eq27}). Namely, for the radial component of the electric
field it follows from (\ref{eq18}) that 
\begin{equation}
E_r(\eta R,z)=-\frac{1}{R}\,\frac{\partial}{\partial \eta}\phi (\eta R,z)= 
-\frac{\sigma}{2\pi\epsilon_0 R}\int\limits_0^\pi\frac{r_1\left (\frac{
\partial r_1}{\partial \eta}\right)_\theta}{\sqrt{r_1^2+z^2}}\,d\theta.
\label{eq29}
\end{equation}
In this expression, the partial derivative $\left (\frac{\partial r_1}
{\partial \eta}\right)_\theta$ should be calculated with constant $\theta$.
On the other hand, it is clear from the Fig.\ref{fig3} that
\begin{equation}
r_1^2=R^2(1+\eta^2-2\eta\cos{2\psi}),
\label{eq30}
\end{equation}
where $\psi$, as equation (\ref{eq19}) indicates, is a function of both $\eta$ 
and $\theta$. Therefore, 
\begin{equation}
r_1\left (\frac{\partial r_1}{\partial \eta}\right)_\theta=r_1\left [
\left(\frac{\partial r_1}{\partial \eta}\right)_\psi+\left (\frac{\partial 
r_1}{\partial \psi}\right)_\eta\left(\frac{\partial \psi}{\partial \eta}
\right)_
\theta\right ].
\label{eq31}
\end{equation}
Differentiating (\ref{eq30}), we get for partial derivatives
\begin{equation}
r_1\left(\frac{\partial r_1}{\partial \eta}\right)_\psi=
R^2(\eta-\cos{2\psi}),\;\; r_1\left(\frac{\partial r_1}{\partial \psi}
\right)_\eta=2\eta R^2\sin{2\psi}.
\label{eq32}
\end{equation}
While it follows from (\ref{eq19}) that
\begin{equation}
\tan{\theta}\left [-2\sin{2\psi}\,\left(\frac{\partial \psi}{\partial \eta}
\right )_\theta -1\right ]=2\cos{2\psi}\,\left(\frac{\partial \psi}{\partial 
\eta}\right )_\theta,
\label{eq33}
\end{equation}
and, therefore,
\begin{equation}
\left(\frac{\partial \psi}{\partial \eta}\right )_\theta=-\frac{\tan{\theta}}
{2(\cos{2\psi}+\tan{\theta}\,\sin{2\psi})}=-\frac{\sin{2\psi}}{2(1-\eta
\cos{2\psi})}.
\label{eq34}
\end{equation}
Substituting (\ref{eq32}) and (\ref{eq34}) into (\ref{eq31}), we get
\begin{equation}
r_1\left (\frac{\partial r_1}{\partial \eta}\right)_\theta=
-\frac{R^2\cos{2\psi}\,(1+\eta^2-2\eta\cos{2\psi})}{1-\eta\cos{2\psi}}.
\label{eq35}
\end{equation}
Combined with (\ref{eq20}) and (\ref{eq29}), this result implies 
\begin{equation}
\fl E_r=\frac{\sigma R}{\pi\epsilon_0}\int\limits_0^{\pi/2}
\frac{\cos{2\psi}}{\sqrt{r_1^2+z^2}}\,d\psi=-\frac{\sigma}{\pi\epsilon_0
\sqrt{(1+\eta)^2+z^2/R^2}}\int\limits_0^{\pi/2}\frac{1-
2\sin^2{\phi}}{\sqrt{1-k^2\sin^2{\phi}}}\,d\phi.
\label{eq36}
\end{equation}
But $\sin^2{\phi}=[1-(1-k^2\sin^2{\phi})]/k^2$ and, respectively, (\ref{eq36}) 
can be immediately expressed in terms of complete elliptic integrals of the 
first and second kinds (in agreement with \cite{14}):
\begin{equation}
E_r=-\frac{\sigma}{\pi\epsilon_0\sqrt{(1+\eta)^2+z^2/R^2}}\left [\left(1-
\frac{2}{k^2}\right)K(k)+\frac{2}{k^2}\,E(k)\right].
\label{eq37}
\end{equation}
The combination of complete elliptic integrals that appear in (\ref{eq37})
can be expressed through the Legendre function of the second kind and 
half-integral order $Q_{1/2}$ (the so called toroidal function of zeroth order
--- see \cite{19} and references therein). Indeed, we have \cite{15,20}
\begin{equation}
Q_{1/2}\left(\frac{2}{k^2}-1\right)=\frac{2-k^2}{k}\,K(k)-\frac{2}{k}\,E(k).
\label{eq38}
\end{equation}
Therefore (\ref{eq37}) can be rewritten as follows:
\begin{equation}
E_r=\frac{\sigma}{2\pi\epsilon_0\sqrt{\eta}}\;Q_{1/2}\left(\frac{1+\eta^2+
z^2/R^2}{2\eta}\right).
\label{eq39}
\end{equation}
It is not immediately clear that (\ref{eq9}) is the $z\to 0$ limit of 
(\ref{eq37}). However this follows from the identities
\begin{equation}
K(n)=(1+\eta)K(\eta),\;\;\;E(n)=\frac{2}{1+\eta}\,E(\eta)-(1-\eta)K(\eta),
\label{eq40}
\end{equation}
that by themselves can be obtained from  Gauss' Transformation formulas for 
complete elliptic integrals (transformations 164.02 in \cite{18}).
 
Similarly, for the vertical component of the electric field, we obtain after 
some simple manipulations
\begin{equation}
\eqalign{ E_z=-\frac{\partial \phi(\eta R,z)}{\partial z}=-\frac{\sigma}
{2\pi\epsilon_0}\left [\int\limits_0^\pi\frac{z}{\sqrt{r_1^2+z^2}}\,d\theta-
\pi\,\mathrm{sign}(z)\right ]= \cr \fl \frac{\sigma}{2\pi\epsilon_0}\left [
\frac{-z/R}{\sqrt{(1+\eta)^2+z^2/R^2}}\int\limits_0^{\pi/2}\frac{\frac{2}
{1+\eta}-n^2\sin^2{\phi}}{\sqrt{1-k^2\sin^2{\phi}}\;(1-n^2\sin^2{\phi})}\,
d\phi+\pi\,\mathrm{sign}(z)\right ].}
\label{eq41}
\end{equation}
Here
\begin{equation}
\mathrm{sign}(z)=\frac{d|z|}{dz}=\left\{\begin{array}{c} 1\;\mathrm{if}\;
z>0,\\  0\;\mathrm{if}\;z=0,\\ -1\;\mathrm{if}\;\;z<0. \end{array} \right .
\label{eq41A}
\end{equation}
Using $\frac{2}{1+\eta}-n^2\sin^2{\phi}=\frac{1-\eta}{1+\eta}+1-\sin^2{\phi}$,
the integral in (\ref{eq41}) can be immediately expressed in terms of the
complete elliptic integrals, and the final answer is
\begin{equation}
\fl E_z=-\frac{\sigma}{2\pi\epsilon_0}\left [\frac{z/R}{\sqrt{(1+\eta)^2+
z^2/R^2}}\left (K(k)+\frac{1-\eta}{1+\eta}\,\Pi(n^2,k)\right )-\pi\,
\mathrm{sign}(z)\right].
\label{eq42}
\end{equation}
This answer does not look identical to the expression given in \cite{14}.
However it coincides with the result obtained in \cite{21} and it was shown
in \cite{21} that the results of \cite{14} and \cite{21} for $E_z$ are actually
equivalent.

\begin{figure}[htp]
\centering
\includegraphics[height=4cm]{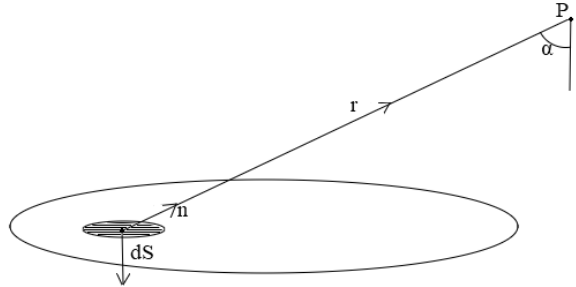}
\caption{Definition of variables in the solid angle formula (\ref{eq43}).
$\vec{n}=\frac{\vec{r}}{r}$ is the unit vector in the direction of $\vec{r}$. }
\label{fig4}
\end{figure}
It follows from the solid angle formula (see Fig.\ref{fig4})
\begin{equation}
\Omega=-\int\frac{\vec{n}\cdot d\vec{S}}{r^2}=\int\frac{\cos{\alpha}}{r^2}
\,dS,
\label{eq43}
\end{equation}
that the $z$-component of the gravitational attraction at $P$ of any plane  
homogeneous lamina with constant surface density is equal to the negative of
surface density times the Newton's constant times the solid angle which the 
lamina subtends at $P$ \cite{22}. In the case of our charged disk, this 
translates into
\begin{equation}
E_z=\frac{\sigma}{4\pi\epsilon_0}\,\Omega,
\label{eq44}
\end{equation}
where $\Omega$ is the solid angle which the disk subtends at $P$. There exist 
independent calculations of $\Omega$ \cite{23,24}. It can be verified that (for
$z>0$) the solid angle $\Omega$ that follows from (\ref{eq42}) and (\ref{eq44})
is consistent with the results of \cite{23,24}.

\section{Electrostatic potential and electric field at points outside the disk 
boundary}
\begin{figure}[htp]
\centering
\includegraphics[height=4cm]{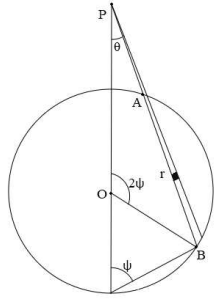}
\caption{Definition of variables when the projection point $P$ lies outside 
the disk boundary.}
\label{fig5}
\end{figure}

If the projection point $P$ lies outside the disk boundary ($\eta>1$), when we 
can arrange only one wedge crossing the disk, as shown in Fig.\ref{fig5}.
When $\theta$ changes from $0$ to $\theta_{max}$, the wedge covers half 
the disk. The limit angle $\theta_{max}$ corresponds the wedge touching the 
disk. Therefore $\sin{\theta_{max}}=1/\eta$. The other symmetric half of the
disk makes an equal contribution to the electrostatic potential, and we can 
write
\begin{equation}
\fl \phi(\eta R,z)=\frac{\sigma}{2\pi\epsilon_0}\int\limits_0^{\theta_{max}}
d\theta \int\limits_{r_1}^{r_2}\frac{r}{\sqrt{r^2+z^2}}\,dr=\frac{\sigma}
{2\pi\epsilon_0}\int\limits_0^{\theta_{max}}\left (\sqrt{r_2^2+z^2}-
\sqrt{r_1^2+z^2}\right )d\theta,
\label{eq45}
\end{equation}
where $r_1=PA$ and $r_2=PB$ (see Fig.\ref{fig5}). Both $r_1$ and $r_2$ obey
the same quadratic equation $R^2=\eta^2R^2+r^2-2\eta R r\cos{\theta}$
with solutions
\begin{equation}
r_1=R\left [\eta\cos{\theta}-\sqrt{1-\eta^2\sin^2{\theta}}\right],\;\;
r_2=R\left[\eta\cos{\theta}+\sqrt{1-\eta^2\sin^2{\theta}}\right ].
\label{eq46}
\end{equation}
In the first integral let's again introduce the angle $\psi$ as defined in 
Fig.\ref{fig5}. Then $r_2^2=R^2(1+\eta^2-2\eta\cos{2\psi})$ and
\begin{equation}
\sqrt{r_2^2+z^2}=R\sqrt{(1+\eta)^2+z^2/R^2}\,\sqrt{1-k^2\sin^2{\phi}},
\label{eq47}
\end{equation}
where $\phi=\pi/2-\psi$ and $k$ was defined in (\ref{eq22}). In addition,
Fig.\ref{fig5} indicates that
\begin{equation}
\tan{\theta}=\frac{R\sin{(\pi-2\psi)}}{\eta R+R\cos{(\pi-2\psi)}}=
\frac{\sin{2\psi}}{\eta-\cos{2\psi}},
\label{eq48}
\end{equation}
and, correspondingly,
\begin{equation}
d\theta=2\frac{\eta\cos{2\psi}-1}{1+\eta^2-2\eta\cos{2\psi}}\,d\psi=
2\frac{1+\eta\cos{2\phi}}{1+\eta^2+2\eta\cos{2\phi}}\,d\phi.
\label{eq49}
\end{equation}
Therefore
\begin{equation}
\fl \int\limits_0^{\theta_{max}}\sqrt{r_2^2+z^2}\,d\theta=
\frac{2R\sqrt{(1+\eta)^2+\frac{z^2}{R^2}}}{(1+\eta)^2}\int\limits_0^
{\frac{\pi}{2}-\psi_m}\frac{(1+\eta-2\eta\sin^2{\phi})(1-k^2\sin^2{\phi})}
{(1-n^2\sin^2{\phi})\sqrt{1-k^2\sin^2{\phi}}}\,d\phi,
\label{eq50}
\end{equation}
where $\psi_m$ corresponds to $\theta_{max}$ and thus satisfies
$\cos{2\psi_m}=\sin{\theta_{max}}=1/\eta$.

\begin{figure}[htp]
\centering
\includegraphics[height=4cm]{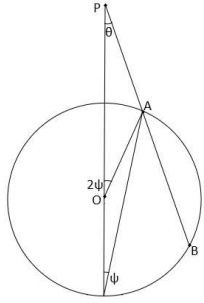}
\caption{Alternative definition of the angle $\psi$.}
\label{fig6}
\end{figure}

In the second integral of (\ref{eq45}), we use an alternative definition of
the angle $\psi$ shown in Fig.\ref{fig6}.
Then $r_1^2=R^2(1+\eta^2-2\eta\cos{2\psi})$, equations (\ref{eq48}) and 
(\ref{eq49}) are still valid, but now $\psi$ changes from $0$ to $\psi_m$,
when $\theta$ changes from $0$ to $\theta_{max}$, and, correspondingly,
$\phi$ changes from $\pi/2$ to $\pi/2-\psi_m$. Therefore
\begin{equation}
\fl \int\limits_0^{\theta_{max}}\sqrt{r_1^2+z^2}\,d\theta=
\frac{2R\sqrt{(1+\eta)^2+\frac{z^2}{R^2}}}{(1+\eta)^2}\int\limits_{\frac{\pi}
{2}}^{\frac{\pi}{2}-\psi_m}\frac{(1+\eta-2\eta\sin^2{\phi})(1-k^2\sin^2{\phi})}
{(1-n^2\sin^2{\phi})\sqrt{1-k^2\sin^2{\phi}}}\,d\phi.
\label{eq51}
\end{equation}  
As we see,the difference of two integrals, (\ref{eq50}) and (\ref{eq51}), 
just gives the right-hand-side of (\ref{eq21}), and, therefore the 
electrostatic potential when $\eta>1$ is still given by (\ref{eq27}), but 
without the last, proportional to $|z|$ term. Introducing Heaviside step
function $H(x)=(1+sign(x))/2$, the final expression for the electrostatic
potential, valid for all values of $\eta$,  takes the form
\begin{equation}
\fl\eqalign{
\;\;\;\;\;\phi(\eta R,z)=\frac{\sigma R}{2\pi\epsilon_0}\left[\frac{1-\eta^2}
{\sqrt{(1+\eta)^2+z^2/R^2}}\,K(k)+\sqrt{(1+\eta)^2+z^2/R^2}\,E(k)+\right .\cr
\left . \;\;\;\;\;
\frac{1-\eta}{1+\eta}\frac{z^2/R^2}{\sqrt{(1+\eta)^2+z^2/R^2}}\,\Pi(n^2,k)-
\pi\,H(1-\eta)\,\frac{|z|}{R}\right].}
\label{eq52}
\end{equation}
The radial electric field is still given by (\ref{eq39}), except that the 
derivative of Heaviside step function will bring an additional term 
proportional to the Dirac delta function:
\begin{equation}
E_r=\frac{\sigma}{2\pi\epsilon_0\sqrt{\eta}}\;Q_{1/2}\left(\frac{1+\eta^2+
z^2/R^2}{2\eta}\right)-\frac{\sigma}{2\epsilon_0}\,\delta(1-\eta)\,\frac{|z|}
{R}.
\label{eq53}
\end{equation}
Finally, the vertical component of the electric field, valid for all values 
of $\eta$, has the form
\begin{equation}
\fl E_z=-\frac{\sigma}{2\pi\epsilon_0}\left [\frac{z/R}{\sqrt{(1+\eta)^2+
z^2/R^2}}\left (K(k)+\frac{1-\eta}{1+\eta}\,\Pi(n^2,k)\right )-\pi\,H(1-\eta)\,
\mathrm{sign}(z)\right].
\label{eq54}
\end{equation}
To use these formulas near the rim of the disk ($\eta\approx 1$), the 
following  asymptotic expansions \cite{18}, valid when $n^2\approx 1$, 
are useful:
\begin{equation}
\fl K(n)\approx \ln{4}-\ln{(1-n^2)},\;\;\;
\Pi(n^2,k)\approx K(k)-\frac{E(k)}{1-k^2}+\frac{\pi(2-k^2-n^2k^2)}
{4\sqrt{(1-n^2)(1-k^2)^3}}.
\label{eq55}
\end{equation}
Then it can be seen from (\ref{eq9}) that the radial electric field has
a logarithmic singularity at the rim of the disc. This singularity is related
to the discontinuity in the surface charge density at $\eta=1$ and can be 
avoided by using a distribution of charge with surface density that falls to 
zero at the rim \cite{15}.

\section{Concluding remarks}
Disc-shaped structures are widespread in astrophysics, their formation being 
associated with the conservation of angular momentum in gas flow around compact
gravitating objects \cite{25}. It is not surprising therefore that the
gravitational counterpart of the considered problem and its generalization to 
the case of non-uniform disks attracted much more attention (see, for example,
\cite{11,12,13,14,15,21,26,27,28,29,30} and references cited therein).

In pedagogical literature, one can find considerations of the gravitational 
field of a massive ring \cite{31,32,33}, and of the electric field of 
a homogeneous ring \cite{19,34,35,36,37}. However, we were able to find only
one paper \cite{38} discussing the gravitational field of a homogeneous
massive disk. 

In \cite{38} the gravitational field of a hypothetical flat Earth was 
considered in the approximation $\eta\ll 1$ \footnote{You will not believe, 
there are many people nowadays who think that the Earth is flat. They even 
hold their own Flat Earth International Conferences \cite{39}. Alas, 
calculations, as in \cite{38}, cannot change their mind --- they do not 
believe in gravity. For every educated person, starting from the ancient times 
of Greek and Roman writers, the sphericity of the Earth was an indisputable  
fact based on solid empirical data \cite{40,41}. The fact that flat Earth  
proponents frantically and ignorantly attack this ancient knowledge is an 
alarming sign of the growing irrationality and decline of faith in the 
scientific method in modern society \cite{39}.}.   
To get the radial electric field in the $\eta\ll 1$ approximation from
(\ref{eq39}), we use the expansion \cite{19,42}
\begin{equation}
\fl Q_{1/2}(\beta)=\frac{\pi}{2(2\beta)^{3/2}}\,\sum\limits_{n=0}^\infty
\frac{(4n+1)!!}{2^{2n}(n+1)!n!}\,\frac{1}{(2\beta)^{2n}}=\frac{\pi}{2(2\beta)
^{3/2}}\,\left (1+\frac{15}{8}\,\frac{1}{(2\beta)^2}\ldots \right),
\label{eq56}
\end{equation}
with $2\beta=(1+\eta^2+z^2/R^2)/\eta$. As a result we get
\begin{equation}
E_r\approx \frac{\sigma\,\eta}{4\epsilon_0(1+z^2/R^2)^{3/2}}\,\left[1+
\frac{3(1-4z^2/R^2)}{8(1+z^2/R^2)^2}\,\eta^2\right ].
\label{eq57}
\end{equation}
For the vertical component of the electric field, we use expansions \cite{18}
\begin{equation}
\eqalign{\fl K(k)=\frac{\pi}{2}\left [1+\frac{1}{4}\,k^2+\frac{9}{64}\,k^4+
\ldots \right ], \; \Pi(n^2,k)=\frac{\pi}{2}\sum\limits_{m=0}^\infty\sum
\limits_{j=0}^m\frac{(2m)!\,(2j)!\,k^{2j}\,n^{2(m-j)}}{4^m\,4^j\,(m!)^2\,
(j!)^2}= \cr  \frac{\pi}{2}\left [1+\frac{1}{2}n^2+\frac{1}{4}k^2+
\frac{3}{8}n^4+\frac{1}{16}k^2n^2+\frac{9}{64}k^4+\ldots \right ],}
\label{eq58}
\end{equation}
and get in the $\eta\ll 1$ approximation (for $z>0$):
\begin{equation}
E_z\approx \frac{\sigma}{4\epsilon_0}\left [2\left (1-\frac{z/R}
{\sqrt{1+z^2/R^2}}\right)-\frac{3z/R}{2\,(1+z^2/R^2)^{\,5/2}}\,\eta^2
\right ].
\label{eq59}
\end{equation}
Under substitution $1/(4\pi\epsilon_0)\to -G$, $G$ being the Newton 
gravitational constant, (\ref{eq57}) and (\ref{eq59}) reproduce the results 
of \cite{38}.

In fact, the problem addressed in this note was resolved a long time ago, much 
earlier than is commonly thought. Conway notes in \cite{43} that the fact
that the potential of a homogeneous disk is expressible in terms of elliptic 
integrals was already known to Weber in 1873 \cite{44}. However, Weber does 
not provide any reference.

The footnote in \cite{12} gives a hint that according to Todhunter \cite{45} 
Giovanni Antonio Amedeo Plana calculated gravitational pulls from the 
ring and from the circular disk. We were able to find Plana's publication 
\cite{46}. In it, Plana indeed gives the gravitational attraction
of a homogeneous circular disk in terms of complete elliptic integrals. 
The corresponding expression for the gravitational potential was given much
later by Arthur Cayley \cite{47}.

It seems that the contributions of Plana and Cayley have now been forgotten,
because all the modern articles that we checked cite \cite{10,13,14} 
as primary sources where the problem of a homogeneous disk was solved.
We think that the upcoming  bicentennial anniversary of the Giovanni 
Plana's paper \cite{46} is a good occasion both to restore the legacy of Plana 
and Cayley, and to make this venerable problem accessible to a wide 
audience of physics students.


\section*{References}
\begin{thebibliography}{99}
\bibitem{1}
Griffiths D J 2013 {\it Introduction to electrodynamics} (New York: Pearson 
Education)

\bibitem{2}
Franklin J 2005 {\it Classical electromagnetism} (New York: Pearson 
Education)

\bibitem{3}
Ohanian H C 1988 {\it Classical electrodynamics} (Boston: Allyn and Bacon)

\bibitem{4}
Vanderlinde J 2004 {\it  Classical electromagnetic theory} (Dordrecht:
Kluwer Academic Publishers)

\bibitem{4A}
Reitz J R, Milford F J 1960 {\it Foundations of electromagnetic theory}
(Reading: Addison-Wesley)

\bibitem{5}
Purcell E M, Morin D J 2013 {\it Electricity and magnetism} (Cambridge:
Cambridge University Press)

\bibitem{6}
Jackson J D 1999 {\it Classical electrodynamics} (New York: John Wiley \& Sons)

\bibitem{6A}
Smythe W R 1989 {\it Static and dynamic electricity} (London: Taylor and 
Francis)

\bibitem{6B}
Greiner W 1998 {\it Classical electrodynamics} (New York: Springer-Verlag)

\bibitem{7}
Zangwill A 2013 {\it Modern electrodynamics} (Cambridge: Cambridge University 
Press)

\bibitem{8}
Schwartz M 1972 {\it Principles of electrodynamics} (New York: McGraw-Hill)

\bibitem{9}
Konopinski E J 1981 {\it Electromagnetic fields and relativistic particles}
(New York: McGraw-Hill)

\bibitem{9A}
Good R H, Nelson T J 1974 {\it Classical theory of electric and magnetic
fields} (New York: Academic Press)

\bibitem{9B}
Panofsky W K H, Phillips M 1962 {\it Classical electricity and magnetism} 
(Reading: Addison-Wesley)

\bibitem{9C}
Stratton J A 2007 {\it Electromagnetic theory} (Hoboken: John Wiley \& Sons)

\bibitem{9D}
Eyges L 2012 {\it The classical electromagnetic field} (New York: Dover)

\bibitem{10}
Durand E 1953 {\it \'{E}lectrostatique. Vol. I. Les distributions} 
(Paris: Masson et Cie)

\bibitem{11}
Duboshin G N 1961 {\it The theory of attraction} (Moscow: Fizmatlit) 
(in Russian)

\bibitem{12}
Kondrat'ev B P 2007 {\it Potential theory: new methods and problems with 
solutions} (Moscow: Mir) (in Russian)

\bibitem{13}
Lass H, Blitzer L 1983 The gravitational potential due to uniform disks and 
rings {\it  Celestial Mech.} {\bf 30} 225-228

\bibitem{14}
Krogh F T, Ng E W, Snyder W V 1982  The gravitational field of a disk
{\it  Celestial Mech.} {\bf 26} 395-405

\bibitem{15}
Conway J T 2002 Analytical solutions for the Newtonian gravitational field 
induced by matter within axisymmetric boundaries {\it Mon. Not. Roy. Astron. 
Soc.} {\bf 316} 540-554

\bibitem{16}
Routh E J 1922 {\it A treatise on analytical statistics, volume 2} (Cambridge:
Cambridge University Press)

\bibitem{17}
Friedberg R 1993 The electrostatics and magnetostatics of a conducting disk
{\it  Am. J. Phys.} {\bf 61} 1084-1096

\bibitem{18}
Byrd P F, Friedman M D 1971 {\it Handbook of Elliptic Integrals
for Engineers and Scientists, Second Edition} (Berlin: Springer-Verlag)

\bibitem{18A}
Abramowitz M,  Stegun I A 1972 {\it Handbook of Mathematical Functions}
(New York: Dover)

\bibitem{19}
Selvaggi J, Salon S, Chari M V K 2007 An application of toroidal functions in 
electrostatics {\it  Am. J. Phys.} {\bf 75} 724-727

\bibitem{20}
Snow C 1954 {\it Formulas for Computing Capacitance and Inductance} 
(Washington: National Bureau of Standards circular 544)

\bibitem{21}
Fukushima T 2010 Precise computation of acceleration due to uniform ring or 
disk {\it  Celest. Mech. Dyn. Astron. } {\bf 108} 339-356

\bibitem{22}
Kellogg O D 1967 {\it Foundations of Potential Theory, Berlin} 
(Berlin: Springer-Verlag)

\bibitem{23}
Paxton F 1959 Solid Angle Calculation for a  Circular Disk {\it Rev. Sci. 
Instrum. } {\bf 30} 254-258

\bibitem{24}
Conway J T 2010 Analytical solution for the solid angle subtended at any point 
by an ellipse via a point source radiation vector potential
{\it Nucl. Instrum. Meth. A} {\bf 613} 17-27 

\bibitem{25}
Regev O, Umurhan O M, Yecko P A 2016 {\it Modern Fluid Dynamics for Physics 
and Astrophysics} (New York: Springer).

\bibitem{26}
Eckhardt D H, Pesta\~{n}a  J L G 2002 Technique for Modeling the Gravitational 
Field of a Galactic Disk {\it ApJ} {\bf 572} L135-L137

\bibitem{27}
Fukushima T 2016 Numerical computation of gravitational field of infinitely 
thin axisymmetric disc with arbitrary surface mass density profile and its 
application to preliminary study of rotation curve of M33 {\it Mon. Not. Roy. 
Astron. Soc.} {\bf 456} 3702-3714

\bibitem{28}
Hur\'{e} J M 2012 A key-formula to compute the gravitational potential of 
inhomogeneous discs in cylindrical coordinates {\it Celest. Mech. Dyn. Astr.}
{\bf 114}  365-385

\bibitem{29}
Nieto M M 2005 Analytic Gravitational-Force Calculations for Models of the 
Kuiper Belt, with Application to the Pioneer Anomaly {\it Phys. Rev.} D 
{\bf 72}, 083004

\bibitem{30}
Pierens A,  Hur\'{e} J M 2004 Rotation Curves of Galactic Disks for Arbitrary 
Surface Density Profiles: A Simple and Efficient Recipe {\it ApJ} {\bf 605} 
179-182

\bibitem{31}
Schumayer D, Hutchinson D A W 2019  Peculiarities in the gravitational field 
of a filamentary ring {\it Am. J. Phys.} {\bf 87} 384-394

\bibitem{32}
Tobin R W, West J 2006 Close orbits about a massive thin ring
{\it Eur. J. Phys.} {\bf 27} 215-223

\bibitem{33}
West J, Dassanayake S, Daniel A  1998 Bound orbits with positive energy
{\it Am. J. Phys.} {\bf 66}, 25-28 

\bibitem{34}
Zypman F R 2006 Off-axis electric field of a charged ring {\it Am. J. Phys.}
{\bf 74} 295-300 

\bibitem{35}
Noh H R 2017 Electrostatic potential of a charged ring: Applications to 
elliptic integral identities {\it J. Korean Phys. Soc.} {\bf 71} 37-41

\bibitem{36}
Datta S 2007 Electric and magnetic fields from a circular coil using elliptic 
integrals {\it Phys. Educ. India} {\bf 24} 203-212

\bibitem{37}
Ciftja O, Babineaux A and Hafeez N 2009 The electrostatic potential of a 
uniformly charged ring {\it Eur. J. Phys.} {\bf 30} 623-627

\bibitem{38}
Kuzii O, Rovenchak A 2019 What the gravitation of a flat Earth would look like 
and why thus the Earth is not actually flat {\it Eur. J. Phys.} 
{\bf 40} 035008

\bibitem{39}
McIntyre L 2019 Calling all physicists {\it Am. J. Phys.} {\bf 87} 694-695

\bibitem{40}
Dreyer J L E 1953 {\it A History of Astronomy from Thales to Kepler}
(New York: Dover Publications)

\bibitem{41}
Russell J B 1991 {\it Inventing the Flat Earth: Columbus and Modern Historians}
(New York: Praeger)

\bibitem{42}
Selvaggi J, Salon S, Kwon O, Chari M V K 2004 Calculating the external magnetic
field from permanent magnets in permanent-magnet motors --- An alternative 
method {\it IEEE Trans. Magnetics} {\bf 40} 3278-3285

\bibitem{43}
Conway J T 2016 Vector potentials for the gravitational interaction of 
extended bodies and laminas with analytical solutions for two disks
{\it  Celest. Mech. Dyn. Astr.} {\bf 125} 161-194

\bibitem{44}
Weber H 1873 Ueber die Besselschen functionen und ihre anwendung auf die 
theorie der elektrischen str\"{o}me {\it J. Reine Angew. Math. } {\bf 75}
75-105

\bibitem{45}
Todhunter I 1873 {\it History of the Mathematical Theories of Attraction and 
the Figure of the Earth} (London: Constable \& Company)

\bibitem{46}
Plana J 1820 Solution de diff\'{e}rents probl\'{e}mes relatifs \'{a} la loi de 
la r\'{e}sultante de l'attraction exerc\'{e}e sur un point mat\'{e}riel par le 
cercle, les couches cylindriques, et quelques autres corps qui en d\'{e}pendent
 par la forme de leurs \'{e}l\'{e}ments {\it M\'{e}morie della Reale Accademia 
delle Scienze in Torino} {\bf 24} 389-450
\url{https://www.biodiversitylibrary.org/item/32409\#page/461/}

\bibitem{47}
Cayley A 1875  On the Potential of the Ellipse and the Circle 
{\it  Proc. Lond. Math. Soc.} {\bf 6}  38-58


\end{thebibliography}
\end{document}